\documentclass[aps,showpacs]{revtex4}
\usepackage{epsf,latexsym}

\begin{document}

\title{Entropy of gravitating systems: scaling laws versus radial profiles}
%\author{Alessandro Pesci\footnotetext{e-mail: pesci@bo.infn.it}}
\author{Alessandro Pesci}
\email{pesci@bo.infn.it}
\affiliation
{INFN, Sezione di Bologna, Via Irnerio 46, I-40126 Bologna, Italy}

\begin{abstract}
Through the consideration of spherically symmetric 
gravitating systems consisting of perfect fluids
with linear equation of state 
constrained to be in a finite volume,
an account is given of the properties of entropy
at conditions in which
it is no longer an extensive quantity
%***
(it does not scale with system's size).
%***
To accomplish this, 
the methods introduced by Oppenheim \cite{Oppen2}
to characterize non-extensivity
are used, suitably generalized
to the case
of gravitating systems subject to an external pressure.
In particular when, far from the 
%***
system's Schwarzschild limit,  
%***
both area scaling for conventional entropy 
and inverse radius law for the temperature 
set in
(i.e. the same properties of the corresponding
black hole thermodynamical quantities),
the entropy profile is found to behave like $1/r$,
being $r$ the area radius inside the system.
In such circumstances thus entropy heavily resides
in internal layers,
in opposition to
what happens when
area scaling
is gained while approaching the Schwarzschild mass,
in which case conventional entropy lies 
at the surface of the system.
The information content of these systems,
even if it globally scales like the area, 
is then stored
in the whole volume,
%***
instead of packed on the boundary.
%***
\end{abstract}


\maketitle

\section{Introduction}
The study of the connection between
entropy and gravity, performed even simply at a classical level,
has deserved in recent years very interesting results.
%***
Examples have been found in which the conventional
%***
(as opposed to black hole) entropy of gravitating
systems scales like their area and the temperature
like $1/R$ (being $R$ their area radius),
that is 
the same as the celebrated laws of black hole entropy
and temperature if area and radius
refer to horizon \cite{hawking,bekenstein}.

Black hole properties are apparent at 
a semi-classical level in which quantum fields
live in a continuous curved spacetime, describing gravity;
these properties moreover, in particular the area
scaling law of black hole entropy, are considered to give hints
or constraints on a fundamental description 
of spacetime itself, possibly at a quantum level,
of a holografic nature \cite{thooft,susskind}.
Now, the fact that purely classical systems 
can exhibit analogous scaling laws in the
corresponding conventional thermodynamical
quantities is intriguing.
The study of these systems
could shed light on what in the area scaling property
of black hole entropy
truly points to a fundamental origin.
 
In \cite{banks} gravitating systems consisting
of perfect fluids with linear equation of state
$p = \kappa \rho$, with $p$ and $\rho$ the pressure and
energy density and $\kappa$ a (non-negative) constant, 
have been considered. 
The authors have found the remarkable property
that for the stiffest case compatible
with causality, namely the $\kappa=1$ case (see \cite{zeldovich}), the equilibrium
configurations can exhibit area scaling of their entropy,
as well $1/R$ scaling of their temperature,
suggesting a possible connection
between holografy and causality.
This recalls
the formulation of the holographic principle,
as given by \cite{bousso}, which has causality built-in.
In \cite{Oppen1},
the entropy of a spherically symmetric gravitating system
in equilibrium has been shown to become
area scaling as the radius approaches
the Schwarzschild value; in this limit the total entropy
of the body is accounted for only by the entropy of the more external
spherical shells.

In general when gravity enters the game,
extensivity of entropy is lost
even if, by the equivalence principle,
it is mantained locally.
%***
In \cite{Oppen2} a method has been introduced
to study the degree of non-extensivity
of entropy (its scaling behaviour with system's size) 
for systems with long range
interactions
(other approaches to non-extensivity, 
trying to characterize and understand it
outside standard thermodynamics, 
use Tsallis entropy \cite{Tsallis}; 
see also \cite{Renyi}). 
%***
Applying this method to a particular
system consisting of a spherically symmetric
distribution of perfect fluid with constant
density, it appears that at increasing $M/R$
($M$ is the mass, or mass-energy, as spelled out in \cite{Wheeler},
known also as ADM mass)
entropy starts going towards area scaling, with
an increasing contribution from external layers
(the limit $R=2M$ is however not reachable
as equilibrium is possible only
if $R \geq (9/4)M$ (see, for instance, \cite{Wheeler})),
in agreement with what expected from \cite{Oppen1}.
The idea in the present paper is 
to go back to the systems \cite{banks},
known to reach conditions with area scaling
of entropy, 
and to investigate,
through a generalization of the method introduced in \cite{Oppen2},
the dependence of non-extensivity on $M/R$
as well to study
the evolution of the entropy profile.

%***
The paper is organized as follows.
In section II we examine the equilibrium configurations
of the fluids under consideration
first under conditions in which they are unconstrained
and next when they are instead constrained to be in a fixed volume.
For these latter conditions
in section III we try to characterize the degree of non-extensivity of the systems
following the approach \cite{Oppen2}, trying to extend it to cover the case
in which an external pressure is applied, i.e. our case.
A global temperature is introduced and
an expression is found that for each $\kappa$
gives the degree of non-extensivity of the entropy of the system
for assigned (central) energy density and size.
In section IV first we use this result to investigate the scaling
laws for entropy; 
next the scaling laws for global temperature
are also derived and, finally,
the entropy radial profiles are considered,
in their relation with the scaling laws of entropy,
in particular for the most interesting case $\kappa = 1$.
We conclude with a summary of our results in section V.
%***

\section{Equilibrium configurations for linear perfect fluids}

As in \cite{banks}, let us consider
a perfect fluid at thermodynamic equilibrium
with linear equation of state

\begin{eqnarray}\label{state}
p = \kappa \rho,
\end{eqnarray}
where $\rho$ and $p$ are respectively the energy density
and pressure in the rest frame of the fluid element;
$\kappa$ is a constant with
$0 \leq \kappa \leq 1$
where, as said, $\kappa = 1$ corresponds to the
stiffest conditions compatible with causality.
The stress-energy tensor is

\begin{eqnarray}\label{stress-energy}
T_{\alpha\beta} = (\rho + p) u_{\alpha} u_{\beta} + p g_{\alpha\beta},
\end{eqnarray}
where $u_{\alpha}$ is the fluid element 4-velocity and
$g_{\alpha\beta}$ is the metric tensor.
In the case of static equilibrium configurations
(necessarily with spherical symmetry),
from (\ref{state}) and (\ref{stress-energy})
Einstein field equations in Schwarzschild coordinates
%***
can be written 
%***

\begin{eqnarray}\label{einstein1}
\nonumber
& & \frac{h^\prime}{r h^{2}} + \frac{h-1}{h r^{2}} = 8 \pi \rho \ \ \ \ 
\qquad \qquad \qquad \qquad \quad \qquad \qquad (tt) \\
& & \frac{f^\prime}{r f h} - \frac{h-1}{h r^{2}} = 8 \pi p \ \ \ 
\qquad \qquad \qquad \qquad \quad \qquad \qquad (rr) \\
\nonumber
& & \frac{f^\prime}{r f h} - \frac{h^\prime}{r h^{2}} +
\frac{1}{\sqrt{f h}} \left( \frac{f^\prime}{\sqrt{f h}} \right)^\prime = 
16 \pi p \ \ \  \qquad \qquad (\theta\theta)=(\phi\phi),
\end{eqnarray}
where the derivative symbol means derivative
with respect to radial coordinate (area radius) $r$ and
where the metric is

\begin{eqnarray}\label{metric}
ds^2 = -f(r) \ dt^2 + h(r) \ dr^2 + r^2 \ d\Omega^2,
\end{eqnarray}
with $f$ and $h$ two non-negative functions of $r$, to be determined.

%***
The properties of the
solutions to these equations are known 
since a long time \cite{Chandra}.
The equilibrium configurations
are completely determined by the equation of state (\ref{state});
they can be derived solving directly the
Tolman-Oppenheimer-Volkoff equation (\cite{Tolman1939,OV} and \cite{Chandra}).
Aiming however at obtaining entropy profiles,
we want to stress the dependence on temperature.
We choose then to solve equations (\ref{einstein1})
inserting explicitly
this dependence since start.
%***
For static gravitational fields, Tolman relation \cite{Tolman1930}
prescribes that $T \sqrt{-g_{00}}$ be constant throughout the system,
where $T$ is the temperature in the rest frame of the fluid element.
This means for us that 

\begin{eqnarray}\label{tolman}
T(r) \sqrt{f(r)} = const.
\end{eqnarray}
For our fluid, thermodynamics gives
the following scaling laws \cite{banks2,banks3}

\begin{eqnarray}\label{banks_eq_rho}
& & \rho \propto T^{(\kappa+1)/\kappa}, \\
\label{banks_eq_s}
& & s \propto T^{1/\kappa},
\end{eqnarray}
with $s$ the entropy density.
The proportionality constants 
depend on the particular realization of the fluid.
Applying these scaling laws locally throughout
our system (equivalence principle),
from equation (\ref{tolman}) we have thus

\begin{eqnarray}\label{C}
\rho(r) f^{\frac{\kappa+1}{2\kappa}}(r) = const \equiv \frac{C}{8 \pi},
\end{eqnarray}
with $C$ some constant with the dimensions
of energy density (length$^{-2}$ in geometrized units).
Note that while the value of $C$ depends
on the chosen unit for length,
the quantities $\rho/C$ and $r \sqrt{C}$ are dimensionless.

Making use of (\ref{C}), 
the first and the second of equations (\ref{einstein1}) become

\begin{eqnarray}\label{einstein2_1}
& & \frac{C}{f^{\frac{\kappa+1}{2\kappa}}} = \frac{h^\prime}{r h^2} + \frac{h-1}{h r^2} \\
\label{einstein2_2}
& & \kappa \frac{C}{f^{\frac{\kappa+1}{2\kappa}}} = \frac{f^\prime}{r f h} - \frac{h-1}{h r^2},
\end{eqnarray}
while the third, as can be verified
(and as noticed in \cite{photonstars}, in the case of a photon
gas), 
is always identically satisfied
if equations (\ref{einstein2_1}-\ref{einstein2_2}) are satisfied,
as a confirmation of Tolman result.

Before we start to solve these coupled differential equations,
let us study some general characteristics which 
the solutions $f(r)$ and $h(r)$
should satisfy. 
$h(r)$ can be put in the form

\begin{eqnarray}\label{hm}
h(r) = \frac{1}{1-2 m(r)/r},
\end {eqnarray}
(see for example \cite{Wheeler}) with $m(r)$
given by

\begin{eqnarray}\label{ADM}
m(r) = \int_0^r{4 \pi {\tilde r}^2 \rho({\tilde r}) d{\tilde r}} =
\frac{C}{2} \int_0^r{\frac{{\tilde r}^2}{f^g({\tilde r})} d{\tilde r}}, 
\end{eqnarray}
that is, $m(r)$ is the ADM mass inside $r$. 
Here we have introduced the function

\begin{eqnarray}\label{g}
g(\kappa) = \frac{\kappa+1}{2\kappa},
\end{eqnarray}
so that $g(\kappa)$ is monotonically
decreasing 
in the relevant range
$0 < \kappa \leq 1$,
with $g(\kappa) \geq 1$ 
and $g(\kappa)\rightarrow\infty$
for $\kappa$ approaching 0.
The cases of perfect fluids with $\kappa=0$ (dust),
$\kappa=1/3$ (blackbody radiation)
and $\kappa=1$ (stiff matter)
correspond respectively to
$g = \infty$, $g = 2$ and $g = 1$.
In the following we will often refer 
to $g$ instead of $\kappa$.

For each allowed equilibrium configuration,
in the limit $r \rightarrow 0$

\begin{eqnarray}\label{mnear0} 
m(r) \simeq \frac{4}{3} \pi r^3 \rho(0),
\end{eqnarray}
so that $m(r)/r \rightarrow 0$. This implies $h(0) = 1$ for each solution
with $\rho(0)$ finite.
As regards $f$, from (\ref{C}) we have

\begin{eqnarray}\label{fbegin}
\rho(0) f^g(0) = \frac{C}{8 \pi},
\end{eqnarray}
so that finite $\rho(0)$ implies a non zero $f(0)$.   
Taking the derivative of (\ref{hm}) and using (\ref{mnear0}) and (\ref{fbegin}),
in the limit $r \rightarrow 0$ we have $h^\prime(r) \simeq \frac{16}{3} \pi \rho(0) r$
and $f^\prime(r) \simeq r f(0) \frac{16}{3} \pi \frac{g+1}{2g-1} \rho(0)$,
so that $h^\prime(0) = 0$ and $f^\prime(0) = 0$.

We want now to solve equations (\ref{einstein2_1}-\ref{einstein2_2}).
We can write equation (\ref{einstein2_2}) as

\begin{eqnarray}\label{h15}
h = (2g - 1) \frac{f^{g-1} (f + r f^\prime)}{C r^2 + (2g-1) f^g}
\end{eqnarray} 
and, substituting this expression of $h$ in equation (\ref{einstein2_1}),
we obtain 

\begin{eqnarray}\label{fderivata2}
f^{\prime\prime} = \frac{2 (g+1) C r f^2 + 3 g C r^2 f f^\prime - 2 (2g-1) f^{g+1} f^\prime 
+ g C r^3 {f^\prime}^2}{C r^3 f + (2g-1) r f^{g+1}}
\end{eqnarray}
so that equations (\ref{h15}-\ref{fderivata2}) are equivalent 
to the starting equations (\ref{einstein2_1}-\ref{einstein2_2}).
Using, in addition, the previous results about the behaviour
of $f$ near the origin, 
the problem is thus reduced to find the solutions $f$
of the second order equation (\ref{fderivata2})
with initial conditions

\begin{eqnarray}\label{condiniz}
& f(0) = \Big(\frac{C}{8 \pi \rho(0)}\Big)^{1/g}; & \qquad \qquad f^\prime(0) = 0.
\end{eqnarray}

For assigned $g$, for each choice of the pair $C$, $\rho(0)$
we have a solution $f(r)$.
Note however from equation (\ref{C}) that
physical configurations $\rho = \rho(r)$ 
depend only on the ratio $C/f^g$ and not on
$C$ and $f$ separately. 
This implies that for any chosen $\rho(0)$,
the (infinite number of) solutions $f$ that we find
when $C$ is varied, correspond always to the same physical
configuration $\rho = \rho(r)$.
For each $g$, to fix $\rho(0)$ corresponds to uniquely fix
the physical configuration, irrespective of the separate values
of $C$ and $f$. 
Without loss of generality we can then study
the solutions to equation (\ref{fderivata2}) for
a single arbitrary choice of the value of $C$ ($\not= 0$);
let us put for simplicity $C = 1$.
All possible static configurations of our system
are then in 1-1 correspondence with all the solutions
$f$ of equation

\begin{eqnarray}\label{fderivata2bis}
f^{\prime\prime} = \frac{2 (g+1) r f^2 + 3 g r^2 f f^\prime - 2 (2g-1) f^{g+1} f^\prime 
+ g r^3 {f^\prime}^2}{r^3 f + (2g-1) r f^{g+1}}
\end{eqnarray}
with initial conditions

\begin{eqnarray}\label{condinizbis}
& f(0) = \Big(\frac{1}{8 \pi \rho(0)}\Big)^{1/g}, & \qquad \qquad f^\prime(0) = 0,
\end{eqnarray}
where $\rho(0)$ is any positive number
and with $h$ given by (\ref{h15}).
Due to (\ref{banks_eq_rho}), each specification of $\rho(0)$
is equivalent to a specification of the central temperature $T_C$.

Some symmetry considerations can now help to understand
how the set of solutions $f$ (and $h$) is arranged.
Consider the transformation

\begin{eqnarray}\label{transformationr}
& & r \rightarrow {\tilde r} = \lambda r \\
\label{transformationf}
& & f \rightarrow {\tilde f} = \lambda^{2/g} f .
\end{eqnarray}
We have

\begin{eqnarray}
\nonumber
{\tilde f}({\tilde r}) = \lambda^{2/g} f({\tilde r}/\lambda)
\end{eqnarray}
or

\begin{eqnarray}
\nonumber
{\tilde f}(r) = \lambda^{2/g} f(r/\lambda)
\end{eqnarray}
and then

\begin{eqnarray}
\nonumber
{\tilde f}^\prime(r) = \lambda^{\frac{2}{g} -1} f^\prime(r/\lambda)
\end{eqnarray}
and

\begin{eqnarray}
\nonumber
{\tilde f}^{\prime\prime}(r) = \lambda^{\frac{2}{g} -2} f^{\prime\prime}(r/\lambda)
\end{eqnarray}
so that, if $f(r)$ is a given solution, ${\tilde f}(r)$ is also a solution,
as can be easily verified going through equation (\ref{fderivata2bis}).
${\tilde f}(r)$ is such that ${\tilde f}(0) = \lambda^{2/g} f(0)$.
On the other hand we know that, 
starting from a given solution $f_1$,
we can obtain all solutions for the assigned $g$
if we change the initial conditions,
$f_1(0) \rightarrow f_{\lambda}(0) = \lambda f_1(0)$,
with $\lambda$ spanning all positive real values.
Thus,
in terms of an assigned solution $f_1(r)$,
a function $f_\lambda$ is a solution if and only if

\begin{eqnarray}\label{allsolutionsf}
f_\lambda(r) = \lambda f_1(r/\lambda^{\frac{g}{2}})
\end{eqnarray}
with $\lambda > 0$.
For the solutions $h$ we obtain

\begin{eqnarray}\label{allsolutionsh}
h_\lambda(r) = h_1(r/\lambda^{\frac{g}{2}}),
\end{eqnarray}
where $h_\lambda(r)$ and $h_1(r)$ corresponds to 
$f_\lambda(r)$ and $f_1(r)$
through (\ref{h15}).
We see that, from (\ref{C}), 
the transformations we are considering
are equivalent to say that equations (\ref{einstein1})
are invariant under the transformation \cite{banks,sorkin,bondi}

\begin{eqnarray}
\nonumber
& & r \rightarrow \lambda r \\
\label{lambdascaling}
& & p \rightarrow p/\lambda^2 \\ 
\nonumber
& & \rho \rightarrow \rho/\lambda^2.
\end{eqnarray}

Summing up, once we know the behaviour of 
solutions $f$ and $h$ for a given particular initial value
$f(0) = a_1$ of $f$, arbitrarily fixed ($>0$),
we can readily get the corresponding behaviours
of the solutions for every other assigned initial value $a_2$.
Let us study for simplicity the case $f(0) = 1$,
that is the solutions to (\ref{fderivata2bis})
with initial conditions

\begin{eqnarray}\label{condinizter}
& f(0) = 1; & \qquad \qquad f^\prime(0) = 0.
\end{eqnarray}

\begin{figure}\leavevmode
\begin{center}
\epsfxsize=8cm
\epsfbox{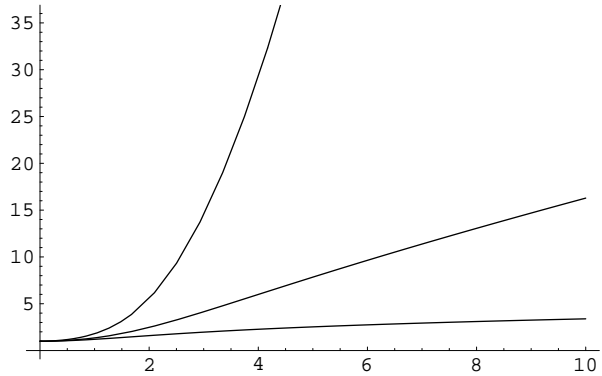}
%***
\caption{$f$ versus $r$ for 
$\kappa=1, \ \frac{1}{3}, \ \frac{1}{9}$ (highest: $\kappa=1$,
lowest: $\kappa=\frac{1}{9}$) 
for initial conditions (\ref{condinizter}), see text.}
%***
\label{f}
\end{center}
\end{figure}

\begin{figure}\leavevmode
\begin{center}
\epsfxsize=8cm
\epsfbox{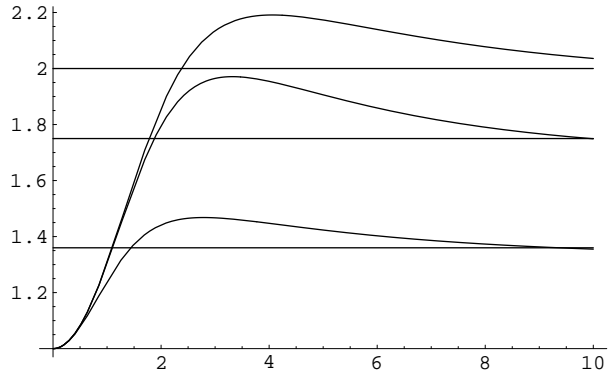}
%***
\caption{$h$ versus $r$ for 
$\kappa=1, \ \frac{1}{3}, \ \frac{1}{9}$ (highest: $\kappa=1$,
lowest: $\kappa=\frac{1}{9}$) 
for initial conditions (\ref{condinizter}), see text.
}
%***
\label{h}
\end{center}
\end{figure}

From numerical integration of equation (\ref{fderivata2bis})
and from equation (\ref{h15}) we obtain
the solutions $f$ and $h$ reported in Figures \ref{f} and \ref{h}
for the cases $g= 1,\ 2,\ 5$ (i.e. $\kappa = 1, \ \frac{1}{3}, \ \frac{1}{3^2}$). 
%%%%%%%%%%%%%%%%%%%%%%%%%%%% figure f %%%%%%%%%%%%%%%%%
%%%%%%%%%%%%%%%%%%%%%%%%%%%%%%%%%%%%%%%%%%%%%%%%%%%%%%%
%%%%%%%%%%%%%%%%%%%%%%%%%%%% figure h %%%%%%%%%%%%%%%%%
%%%%%%%%%%%%%%%%%%%%%%%%%%%%%%%%%%%%%%%%%%%%%%%%%%%%%%%
The structure of the corresponding equilibrium configuration
can be seen in Figures \ref{rho} and \ref{2m_over_r},
that respectively report $\rho$
and $2m/r$, as a function of $r$,
for the same chosen values of $g$.
Our choice $C = 1$ implies that $\rho$ and $r$ are given here
respectively in units of $C$ and $C^{-1/2}$.
From (\ref{C}) and (\ref{allsolutionsf}) the properties
of $\rho$ for every other solution can readily be inferred.
Note that $\rho(r)$ approaches asymptotically 0 as $r \rightarrow \infty$.
This implies that for this configuration and, 
by scaling property (\ref{allsolutionsf}),
for every other solution, the radius of the system is infinite as expected.
%***
These radial profiles agree with the corresponding results reported
in \cite{Chavanis} within an extensive study of the conditions
of general relativistic instability of finite spheres 
consisting of exactly the same fluids we are examining here. 
%***

Let us consider in some detail the functions 
$h$  and $2m/r$ (Figures \ref{h} and \ref{2m_over_r}).
$h$ starts from 1 at $r=0$, 
and $2m/r$ from 0;  
they reach their maxima,
after which they go
towards their own limiting values \cite{photonstars},

\begin{eqnarray}\label{limith}
h(\infty) = 2, \ \frac{7}{4}, \ \frac{34}{25},
\end{eqnarray}
and

\begin{eqnarray}\label{limit}
\left(\frac{2m}{r}\right)_\infty = \frac{1}{2}, \ \frac{3}{7}, \ \frac{9}{34},
\end{eqnarray}
respectively for $g = 1, \ 2, \ 5$,
shown in the Figures, 
with damped oscillations around them,
in which both the amplitudes
of the deviations from the limiting values as well their derivatives
at any order reduce asymptotically to 0.
Analytically both limiting values can easily be verified.
In fact consider for example that from (\ref{hm}) we have

\begin{eqnarray}\label{2m_over_r_formula}
\frac{2m}{r} = \frac{h-1}{h}.
\end{eqnarray}
In equation (\ref{einstein2_1}) we see then that,
at extremal points,
that is when $r$ has values $r^*$ such that
$h^\prime(r^*) = 0$ (that is $(2m(r)/r)^\prime_{r=r^*} = 0$),
we have

\begin{eqnarray}\label{extrema}
\frac{2m(r^*)}{r^*} = \frac{{r^*}^2}{f^g(r^*)}. 
\end{eqnarray}
As extremal values tend to $(2m/r)_\infty$ 
when $r \rightarrow \infty$,
from equation (\ref{einstein2_2}) considered in this limit,
using (\ref{2m_over_r_formula}) we have

\begin{eqnarray}
\nonumber
\frac{1}{2g-1} \left(\frac{2m}{r}\right)_\infty =
\frac{2}{g} \left(1 - \left(\frac{2m}{r}\right)_\infty\right) - \left(\frac{2m}{r}\right)_\infty
\end{eqnarray}
so that

\begin{eqnarray}\label{2m_over_r_far}
\left(\frac{2m}{r}\right)_\infty = \frac{4 - 2/g}{4 - 2/g +2g}
\end{eqnarray}
and
\begin{eqnarray}\label{hinfty}
h(r) = \frac{1}{1-\left(\frac{2m}{r}\right)_\infty} = \frac{4 - 2/g +2g}{2g}
\end{eqnarray}
and thus the values (\ref{limit}),
as well
(\ref{limith}).
Here we made use of the fact that
in the limit $r \rightarrow \infty$,
$(r^2/f^g)^\prime = 0$ implies $f^\prime = \frac{2}{g} f/r$.
Note that equations (\ref{2m_over_r_formula}) 
and (\ref{allsolutionsh})
determine 
$2m/r$ for every solution; we see that $2m/r$ scales
as $h$. 
From scaling property (\ref{allsolutionsh}) and 
from equations (\ref{condinizbis}) and ({\ref{transformationf}),
the asymptotic behaviour of $h$ and $2m/r$ when $r \rightarrow \infty$
implies that in the limit of very large central density $\rho(0)$
or very large $T_C$, $h(r) \simeq const = h(\infty)$ 
and $2m(r)/r \simeq const = \left(\frac{2m}{r}\right)_\infty$,
for every $r$,
except the region of $r$ very near to 0, in which $h$ goes from 1
to $h(\infty)$
and $2m/r$ goes from 0 to $\left(\frac{2m}{r}\right)_\infty$,
with damped oscillations.

As regards $f$,
from (\ref{einstein2_2}) we see that $f^\prime(r) \geq 0, \forall r$
and from (\ref{extrema}) we have that in the limit 
$r \rightarrow \infty$, 

\begin{eqnarray}\label{finfty}
f(r) \simeq \left(\frac{2m}{r}\right)_\infty^{-1/g} 
\ r^{2/g}.
\end{eqnarray}
Due to equation (\ref{allsolutionsf}) this implies

\begin{eqnarray}
\nonumber
& & \lim_{\lambda\to0} f_{\lambda}(r) =  
\lim_{\rho(0)\to\infty} f_{\lambda}(r) =
\lim_{T_C\to\infty} f_{\lambda}(r) = 
\lim_{\lambda\to0} \lambda f(\frac{r}{\lambda^{g/2}}) = \\
\nonumber
& & \lim_{\lambda\to0} \ \lambda \ \left(\frac{2m}{r}\right)_\infty^{-1/g} \ 
\left(\frac{r}{\lambda^{g/2}}\right)^{2/g} =
\left(\frac{2m}{r}\right)_\infty^{-1/g} \ r^{2/g}, \qquad \qquad (r fixed)
\end{eqnarray}
and then

\begin{eqnarray}\label{limitT}
\lim_{\rho(0)\to\infty} \rho(r) =
\lim_{T_C\to\infty} \rho(r) =
\lim_{T_C\to\infty} \frac{1}{8 \pi f_{\lambda}^g(r)} = 
\frac{1}{8 \pi} \ \left(\frac{2m}{r}\right)_\infty \ \frac{1}{r^2}
\qquad \qquad (r \ fixed)
\end{eqnarray}
that is when the central density
is allowed to increase without limit, the density
at each fixed radius $r$ does not increase unlimitedly
but approaches a limiting value proportional to 
$r^{-2}$.
%***
Looking at (\ref{2m_over_r_far}), the proportionality constant here
is such that these limiting conditions correspond to
the analytic (unphysical) solution for the pressure (or energy density)
\cite{misner_zapolsky}
\begin{eqnarray}\label{analytic}
p = \frac{\kappa^2}{2 \pi (1 + 6\kappa + \kappa^2)} \ \frac{1}{r^2},
\end{eqnarray}
in agreement with what stated in \cite{banks, Chandra, Chavanis}.
This analytic solution is represented in Figures \ref{h} and \ref{2m_over_r}
by the horizontal lines; the intercepts of the physical solutions
with the analytic one follow a well known geometrical progression
with ratio depending on $\kappa$
\cite{Chandra, Chavanis}.
%***  

Finally, from scaling properties
(\ref{allsolutionsf}-\ref{allsolutionsh}),
from initial conditions (\ref{condinizbis})
and from the behaviour of $f(r)$, $h(r)$ and $\rho(r)$
near $r=0$, shown in Figures \ref{f}, \ref{h} and \ref{rho},
we have that in the limit of very low central density,
$f(r)$, $h(r)$ and $\rho(r)$ are constant
in a ever increasing range of $r$, 
so that the flat-space limit with constant densities
is recovered.

%%%%%%%%%%%%%%%%%%%%%%%%%%%% figure rho %%%%%%%%%%%%%%%%%
\begin{figure}\leavevmode
\begin{center}
\epsfxsize=8cm
\epsfbox{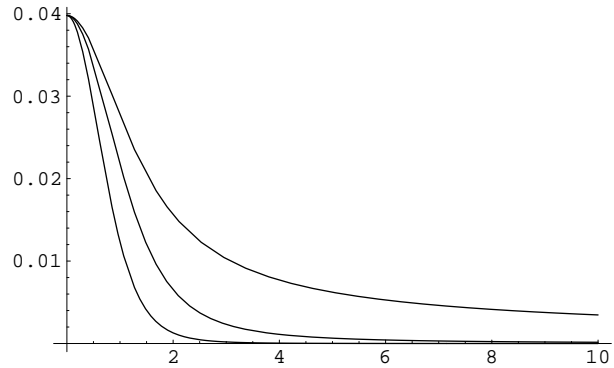}
%***
\caption{$\rho$ versus $r$ for 
$\kappa=1, \ \frac{1}{3}, \ \frac{1}{9}$
(highest: $\kappa=\frac{1}{9}$,
lowest: $\kappa=1$) 
for initial conditions (\ref{condinizter}),
namely $\rho(0)=1/8\pi$, 
see text.}
%***
\label{rho}
\end{center}
\end{figure}
%%%%%%%%%%%%%%%%%%%%%%%%%%%%%%%%%%%%%%%%%%%%%%%%%%%%%%%
%%%%%%%%%%%%%%%%%%%%%%%%%%%% figure 2m_over_r %%%%%%%%%%%%%%%%%
\begin{figure}\leavevmode
\begin{center}
\epsfxsize=8cm
\epsfbox{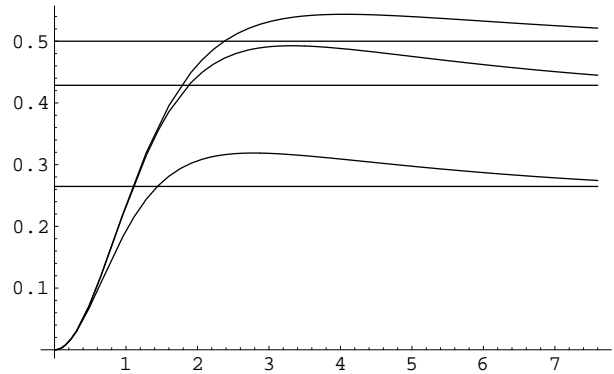}
%***
\caption{$2m/r$ versus $r$ for 
$\kappa=1, \ \frac{1}{3}, \ \frac{1}{9}$ (highest: $\kappa=1$,
lowest: $\kappa=\frac{1}{9}$) 
for the same conditions
as previous Figures.}
%***
\label{2m_over_r}
\end{center}
\end{figure}
%%%%%%%%%%%%%%%%%%%%%%%%%%%%%%%%%%%%%%%%%%%%%%%%%%%%%%%

To obtain now a system with finite size 
we can imagine
to enclose our fluid
inside a perfectly insulating spherical cavity,
with area radius $R$,
and to assume that thermodynamical equilibrium
is reached, that is
the temperature of the walls is equal to
the temperature of fluid at $r = R$.
This will correspond to a certain central
energy density $\rho_C$ for the cavity. 
Due to the local uniqueness of solutions
to equation (\ref{fderivata2bis}) with initial conditions
(\ref{condinizbis}), 
this is equivalent to take a certain (physical) solution
to equation (\ref{fderivata2bis})
(that with $\rho(0) = \rho_C$) and
to cancel all fluid outside $r = R$. 
As regards energy density radial profile $\rho(r)$,
this means that there is a 1-1 correspondence between
all admitted cavities 
filled with linear fluids 
(that is every admitted central densities
as well radii)
and
all possible $\rho(r)$ distributions 
in finite ranges [0,$R$], 
where $\rho(r) = 1/(8 \pi f^g(r))$ and $f$ is solution
to (\ref{fderivata2bis}) with initial conditions (\ref{condinizbis}).
 
As allowed by Birkhoff theorem \cite{Wheeler,Birkhoff},  
if the constraining system is assumed spherically symmetric
and with finite extension,
the Schwarzschild coordinates
can be chosen in such a way that outside a certain radius
$R^*$
exactly the Schwarzschild form of the metric obtains

\begin{eqnarray}\label{schwarzschild}
ds^2 = - ( 1 - 2M^*/r) \ dt^2 + \frac{1}{1-2M^*/r} \ dr^2 
+ r^2 d\Omega^2.
\end{eqnarray}
The parameter $M^*$ here is precisely the mass-energy or ADM mass 
for the whole system, 
that is fluid + constraining system.
Due to our symmetry conditions we can look
at the quantity

\begin{eqnarray}\label{cavityADM}
M = \int_0^R{4 \pi r^2 \rho(r) dr} =
\frac{C}{2} \int_0^R{\frac{r^2}{f^g(r)} dr}, 
\end{eqnarray}
as the mass due to the fluid alone,
that is
the mass would be probed
by Kepler-like experiments with gravitating
bodies far away from the fluid alone.
It represents a meaningful quantity,
proper of the ball of fluid
contained in the cavity,
independent of the characteristics (mass, extension,..)
of the constraining system.
As the metric outside has been explicitly chosen,
the continuity of the metric tensor components $g_{\alpha\beta}$
everywhere, even across boundaries,
fixes, for every assigned $\rho_C$ and $g$,
$f(r)$ $\forall r$.
Due to (\ref{fbegin}) the value of $C$
is fixed also.
In other words, once a single coordinate system is explicitly chosen,
the arbitrariness in the choice of
the constant $C$ in equations 
(\ref{einstein2_1})-(\ref{einstein2_2}) is lost.

Fig. \ref{mass_vs_T} permits to see, for a cavity
with fixed area radius $R$ at equilibrium, 
the dependence of its mass on the
central density or temperature,
for different assigned $g$.
It reports $2M/R$ as a function
of $ z \equiv \rho(0) R^2$.
%%%%%%%%%%%%%%%%%%%%%%%%%%%% figure mass_vs_T %%%%%%%%%%%%%%%%%
\begin{figure}\leavevmode
\begin{center}
\epsfxsize=8cm
\epsfbox{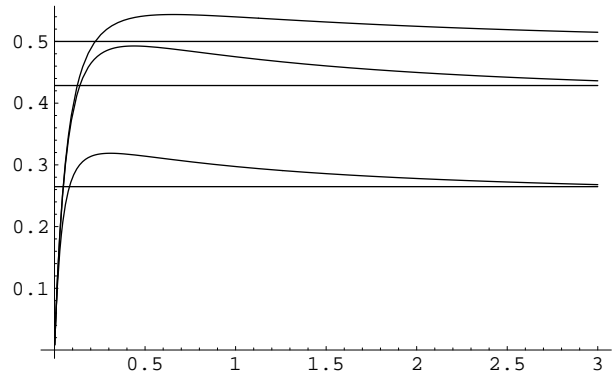}
%***
\caption{$2M/R$ versus $z \equiv \rho(0) R^2$ ($R$ can be thought fixed), for
$\kappa=1, \ \frac{1}{3}, \ \frac{1}{9}$ (highest: $\kappa=1$,
lowest: $\kappa=\frac{1}{9}$).} 
%***
\label{mass_vs_T}
\end{center}
\end{figure}
%%%%%%%%%%%%%%%%%%%%%%%%%%%%%%%%%%%%%%%%%%%%%%%%%%%%%%%
The plot is done in terms of $z$ instead of $\rho(0) \equiv y$
in order to obtain for $2M/R$ an universal shape, 
i.e. not depending on the chosen $R$.
The expression for $z$
indicates how to rescale the abscissa 
to obtain, from the given plot,
the actual relation between $2M/R$ and $y$
for each fixed value of $R$.
The explicit form of the plotted function, 
expressed in terms of the solution
$h_1$ for initial conditions (\ref{condinizter})
(the solution $h$ presented in Fig. \ref{h}), is

\begin{eqnarray}\label{mass-temperature}
\frac{2M(z)}{R} = 
\frac{h_1(\sqrt{8 \pi z}) - 1}{h_1(\sqrt{8 \pi z})}
\end{eqnarray}  
as can be easily derived from equation (\ref{2m_over_r_formula}) and
from the scaling property (\ref{allsolutionsh}).
A maximum $(2M/R)_{max} = 0.544$ is found
(with the extremal point at $z_{extr} = 0.655$),
corresponding to the case $g=1$ ($\kappa =1$); it gives
the maximum fraction of the Schwarzschild mass
a perfect linear fluid at equilibrium confined inside a given radius
can have. The maximum value for the case $g=2$ ($\kappa=1/3$),
namely $(2M/R)_{max} = 0.493$, is in agreement with \cite{sorkin}.
%***
All these results agree with what reported in \cite{Chavanis},
where in addition a prescription is given
to relate $(2M/R)_{max}$ values when $\kappa$ is varied.
%***

In Fig. \ref{mass_vs_T} we see that
once the extremal central energy density or temperature
are overcome, equilibrium configurations are still possible,
at lower values of $M$. We have thus larger central temperatures,
which give rise however to a lower total mass for the system. 
This is by no way a problem for energy conservation, obviously.
The total energy of the system, given by $M$,
contains in fact both the component of proper energy $\rho$
and the (negative) component of gravitational potential 
energy \cite{Wheeler,misner_sharp}.
Evidently, when the extremal temperature is overcome,
the system admits a new equilibrium configuration
with a stronger gravitational energy component.

%***
A detailed discussion 
of the stability of the confined fluids
we are considering here is in \cite{Chavanis}.
According to those results a new mode on instability
occurs whenever a (non trivial)
local extremum in our Fig. \ref{mass_vs_T} is present.
This amounts to say that,
as far the equilibrium of the fluid alone is considered
(neglecting the constraining system \cite{katz}),
starting from a situation of stability at low $z$
(very low energy density and/or very small radius),
the first maximum
marks the transition to instability
and this latter will never desappear for larger values of $z$,
so that stability is for $z \leq z_{extr}$.
This agrees with the results in \cite{sorkin}
corresponding to the case $\kappa = \frac{1}{3}$
and with the general discussion 
of thermodynamic equilibrium and stability 
conditions in \cite{katz2}.
%***

\section{Non-extensivity}
The aim of this section is to characterize the systems under
study in terms of their degree of non-extensivity.
This is done using the methods introduced in \cite{Oppen2}, 
generalizing them to the case of gravitating systems
subject to an external pressure.

The first law of thermodynamics 
as applied to an infinitesimal volume $dV$
of a homogeneus system reads
\begin{eqnarray}\label{first_law}
dE = T \ dS - p \ dV + \mu \ dN,
\end{eqnarray}
where $dE$, $dS$ and $dN$ are respectively
energy, entropy and number of particles in the
volume $dV$,
and $\mu$ the chemical potential.
Integrating this equation to a tiny volume $V$
means to sum over elementary volumes, each possessing
always the same values of the intensive quantities.
If the quantities on which we integrate are extensive,
that is if they scale with size, integration gives

\begin{eqnarray}\label{gibbs-duhem}
E = T S -p V +\mu N,
\end{eqnarray}
so that this relation (known as Gibbs-Duhem relation)
requires and manifests
extensivity. 
This can conveniently be expressed as
$1 = \frac{T S}{E}$ or $1 = \frac{T S}{E + pV}$
or $1 = \frac{T S}{E +p V - \mu N}$ if, respectively,
no $p$ and $\mu$ terms,
$p$ terms or 
$p$ and $\mu$ terms
are chosen to be included in the expression of $S$. 
%***
In particular, without $p$ and $\mu$ terms in $S$, 
a linear relation
\begin{eqnarray}\label{linearity}
S(\lambda E) = \lambda S(E)
\end{eqnarray}
is implied, where the variations of $E$ and $S$
are intended with intensive quantities fixed.
%***
This is certainly the case of most thermodynamic systems
also at macroscopic level (our systems included),
when the volume $V$ is the whole volume of the system at thermodynamic equilibrium,
provided gravitational interaction or, in general,
long range interactions can be neglected 
(as often assumed, at least implicitly, in thermodynamics) \cite{Oppen2}.
In this case the intensive quantities are intended to have values 
$T_0$, $p_0$ and $\mu_0$ characterizing globally
the system.

When however
gravitational effects become no longer negligible
one can expect that relation (\ref{gibbs-duhem}),
albeit holding true locally (equivalence principle),
be modified at a global level
(with a suitable definition of the intensive quantities for this case),
revealing thus non-extensivity.
Correspondingly, relation (\ref{linearity}) should
be modified to

%***
\begin{eqnarray}\label{non-linearity}
S(\lambda M) = \lambda^a \ S(M),
\end{eqnarray}%***
where $a$ is a constant $\neq 1$
and we have indicated with $M$ 
the total energy of the system.
Entropy is no longer extensive
with energy, but related to it
by an Euler relation of homogeneity $a \neq 1$.

Looking at the analogy between the properties of local temperature
in a lattice model with long range
spin-spin interactions and 
the local temperature behaviour for gravitating systems
as given by Tolman relation,
the suggestion has been made in \cite{Oppen2}
to define a global temperature $T_0$ for gravitating systems,
to be identified with the temperature $T_\infty$
that would be measured by an observer far away
(in an asymptotically flat spacetime).
Given this definition,
thermodynamic consistency demands that
for gravitating systems in self-equilibrium,
or, in general, for closed systems not influenced 
mechanically by any external agent,
the equation

\begin{eqnarray}\label{euler}
S = \alpha \frac{M}{T_0}
\end{eqnarray}
be an Euler relation of homogeneity $1/\alpha$ in $M$,
provided $S$ here is intended as total entropy deprived
of its component dependent on $\mu$ (if present).
In fact the first law (\ref{first_law}) implies
$1/T_0 =  \partial{S}/\partial{M}$.
In other terms, putting $\gamma \equiv 1/a$, 
we obtain
$\gamma = \alpha$ or

\begin{eqnarray}\label{gamma}
\gamma = \frac{T_0 S}{M},
\end{eqnarray}
so that any $\gamma \neq 1$ signals and measures
non-extensivity.

Black holes obey
this relation,
as can be verified through
the explicit expressions of black hole temperature and entropy
or, more directly, through the Smarr relation \cite{smarr}

\begin{eqnarray}\label{smarr}
S = \frac{M}{2 T_0}
\end{eqnarray}
(see \cite{Oppen2,Oppen1})
where we see $\gamma = 1/2$,
so that correspondingly $S(M) \propto M^2$.

Analogously (and intriguingly) for any classical gravitating system
in self-equilibrium,
approaching its own Schwarzschild radius,
(conventional) entropy satisfies 
this same relation (\ref{smarr}) 
giving thus again $\gamma = 1/2$ \cite{Oppen1}.
In \cite{Oppen2}, equation (\ref{gamma})
has been applied
to a gravitating system in self-equilibrium 
consisting of a perfect fluid at constant density
and the variation of $\gamma$ with $M/R$
(being $R$ the radius of the system) has been investigated.

Attempting to extend the above analysis to our case,
we face here the problem that,
contrary to the cited cases,
our systems are constrained to be in the assigned volume
by an external pressure $p_{ext}$ applied at $r=R$, 
so that they cannot be considered mechanically insulated.
Applying however
the same reasoning that leads
to equations (\ref{euler}-\ref{gamma}),
we see that these equations still hold true
also in case an external pressure is present,
provided $S$ is intended as total entropy
deprived now also of the component due
to a $p_{ext} \neq 0$.
That is we have

\begin{eqnarray}\label{recast}
\gamma =
\frac{T_0 S_0}{M},
\end{eqnarray}
where with $S_0$
we intend the component of $S$ obtained assuming $p_{ext} = 0$ (in addition to $\mu=0$).

Let us now evaluate total entropy for our systems.
The Gibbs-Duhem relation (\ref{gibbs-duhem}),
as applied locally, expressed in specific form

\begin{eqnarray}\label{gibbs-duhem-diff}
\rho = T s - p + \mu n
\end{eqnarray}
gives

\begin{eqnarray}\label{entropy_local}
s = \frac{1}{T} \ (\rho + p - \mu n).
\end{eqnarray}
%***
From this, considering
that to deprive $s$ of the component of $p$ 
due to the presence of a $p_{ext} \neq 0$ means
to quit the whole $p$
(i.e. $p_{ext} = 0$ implies $p = 0$, 
for the fluids under study if, as at present, they are constrained
to be in a finite volume),
%***
we have 

\begin{eqnarray}\nonumber
S_0(y) = 
\int_0^R 4 \pi r^2 \frac{\rho}{T} \sqrt{h} \ dr =
b^{1/2g} \int_0^R 4 \pi r^2 \left(\frac{1}{8\pi}\right)^\frac{2g-1}{2g}
\left(\frac{1}{f(r)}\right)^\frac{2g-1}{2} \sqrt{h(r)} \ dr
\end{eqnarray}
where $y = \rho(0)$ and
$f$, $h$ are determined by this initial condition with $C=1$
(and with assigned $g$)
and use of equations (\ref{banks_eq_rho}-\ref{C})
is made, with (\ref{banks_eq_rho}) in the form

\begin{eqnarray}\label{b}
\rho = b \ T^{2g},
\end{eqnarray}
with $b$ a constant.
We further obtain

\begin{eqnarray}\label{S1}
S_0(y) =
b^\frac{1}{2g} y^\frac{2g-1}{2g} 4\pi
\int_0^R r^2 \left(\frac{1}{f_1(\sqrt{8\pi y} \ r)}\right)^\frac{2g-1}{2} 
h_1(\sqrt{8\pi y} \ r)^{1/2} \ dr,
\end{eqnarray}
%***
where the expression for $S_0(y)$ depends only on the solutions
%***
for the initial conditions $f(0)=1$;
here use of the equation

\begin{eqnarray}\label{f1}
f(r) = \frac{1}{(8\pi y)^{1/g}} f_1(\sqrt{8\pi y} \ r)
\end{eqnarray}
is made, relation that can readily be obtained
from scaling property (\ref{allsolutionsf}).
Expressing the integration in (\ref{S1}) in terms of the variable
${\tilde z} \equiv y r^2$ we finally obtain

\begin{eqnarray}\label{Sz}
S_0(z) = b^\frac{1}{2g} y^{-\frac{1}{2g}-\frac{1}{2}} 2\pi
\int_0^z \sqrt{\tilde z} \left(\frac{1}{f_1(\sqrt{8\pi {\tilde z}})}\right)^\frac{2g-1}{2}
h_1(\sqrt{8\pi {\tilde z}})^{1/2} \ d{\tilde z},
\end{eqnarray}
where $z = y \ R^2$ (as in (\ref{mass-temperature})).

From equation (\ref{entropy_local}) we see
that $s$ is given by three components,
each in principle with its own radial profile.
For our systems however these radial profiles are the same.
In fact from equations (\ref{banks_eq_rho}) and (\ref{banks_eq_s}),
we see that $s$, $\rho/T$ and $p/T$ terms 
depend only on $T$, with a same dependence, 
so that the same must hold
for the $\mu n/T$ term also.
This assures the $\mu n/T$ term, if present,
has the same radial profile as the $\rho/T$ term.
Total entropy
with all the components included, 
is then proportional to $S_0$ so that the two entropies 
have a same scaling law with $M$.

As regards global temperature $T_0$,
from Tolman relation (\ref{tolman}) and
from (\ref{b}) and (\ref{C})
we have

\begin{eqnarray}\label{T0C}
T_0 = T_C \sqrt{f(0)} = b^{-1/2g} (8\pi)^{-1/2g} C^{1/2g},
\end{eqnarray}
where, as said, $T_0$ is the temperature
measured by a far away observer.
We see that $T_0$ is sensitive to the value of $C$,
that on its own should depend on the mass-energy distribution
of the whole system, constraining system included. 
Global temperature however is,
like energy $M$ or entropy $S$,
a thermodynamical parameter
that, as such, must be characteristic of the chosen system
it describes, irrespective of its environment.
Moreover, if the constraining system is, as we suppose,
in thermal equilibrium with the fluid contained in it,
a same far away temperature $T_0$ must coherently be measured,
independently of the characteristics (density, size, ..)
of the constraining system.
From (\ref{T0C}) this implies that, 
once the asymptotic flatness of the metric is required,
the actual value of $C$ is fixed,
irrespectively of the characteristics of the constraining
system. 
We are thus led to compute $T_0$ as the far away temperature
evaluated as if the constraining system would be absent.
From the continuity of the metric tensor component
on the boundary of the cavity we have then

\begin{eqnarray}\label{boundary}
f(R) = 1 - 2M/R,
\end{eqnarray}
so that from (\ref{C})

\begin{eqnarray}
\frac{C}{(1 - 2M/R)^g} = \frac{1}{f(R)^g},
\end{eqnarray}
where, as above, $f$ is the solution with
$C=1$ and $y=\rho(0)$.
From this equation and from
(\ref{mass-temperature}) and (\ref{f1})
we obtain the following expression for $C$

\begin{eqnarray}
C = 8\pi y \ \frac{1}{[h_1(\sqrt{8\pi z}) f_1(\sqrt{8\pi z})]^g},
\end{eqnarray}
so that $T_0$ is

\begin{eqnarray}\label{T0}
T_0 = b^{-\frac{1}{2g}} y^\frac{1}{2g} \ \frac{1}{[h_1(\sqrt{8\pi z}) f_1(\sqrt{8\pi z})]^{1/2}}.
\end{eqnarray} 

From this expression for $T_0$ and from the expression (\ref{Sz}) for entropy $S$,
$\gamma$ as given by (\ref{recast})
can finally be written.
Noting that $M$, 
given by equation (\ref{mass-temperature}),
can also be written as

\begin{eqnarray}\nonumber
M = 2\pi y^{-\frac{1}{2}} \ \int_0^z\sqrt{\tilde z} f_1(\sqrt{8\pi {\tilde z}})^{-g} \ d{\tilde z}
\end{eqnarray}
(from (\ref{cavityADM}),
once the integration variable is changed from $r$ to ${\tilde z} = y \ r^2$),
for $\gamma$ the expression

\begin{eqnarray}\label{gammaz}
\gamma(z) = 
\frac{T_0 S_0}{M} =
[h_1(\sqrt{8\pi z}) f_1(\sqrt{8\pi z})]^{-1/2}
\cdot
\frac{\int_0^z \sqrt{\tilde z}f_1(\sqrt{8\pi {\tilde z}})^{-g} 
[h_1(\sqrt{8\pi {\tilde z}}) f_1(\sqrt{8\pi {\tilde z}})]^{1/2} \ d{\tilde z}}
{\int_0^z \sqrt{\tilde z}f_1(\sqrt{8\pi {\tilde z}})^{-g} \ d{\tilde z}}
\end{eqnarray}
is found.

\section{Scaling laws and radial profiles}

Fig.\ref{gam} plots $\gamma(z)$ as given by equation (\ref{gammaz}).
%%%%%%%%%%%%%%%%%%%%%%%%%%%% figure gam %%%%%%%%%%%%%%%%%
\begin{figure}\leavevmode
\begin{center}
\epsfxsize=8cm
\epsfbox{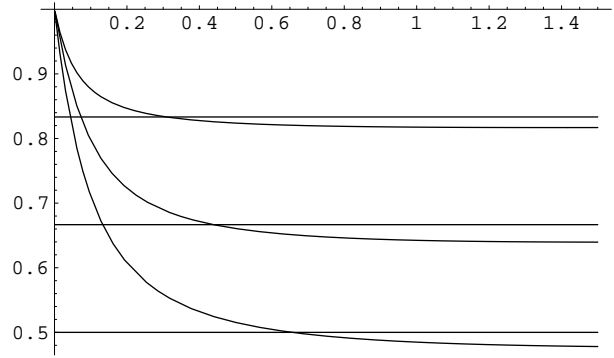}
%***
\caption{$\gamma$ versus $z \equiv \rho(0) R^2$ for 
$\kappa=1, \ \frac{1}{3}, \ \frac{1}{9}$ (highest: $\kappa=\frac{1}{9}$,
lowest: $\kappa=1$).}
%***
\label{gam}
\end{center}
\end{figure}
%%%%%%%%%%%%%%%%%%%%%%%%%%%%%%%%%%%%%%%%%%%%%%%%%%%%%%%
At each $z$ value, we have $S(M) \propto M^{1/\gamma}$ with $\gamma$
given by the plot.
We see that in the limit $z \rightarrow 0$
(low densities and/or small radii),
$\gamma \rightarrow 1$, 
so that $S(M) \propto M$ (and $S(R) \propto R^3$, from (\ref{cavityADM})) 
and extensivity obtains.
This can analytically be seen from (\ref{gammaz}),
as in this limit from (\ref{condinizter}) and (\ref{h15})
we have $f,h \simeq 1$ and $\gamma$ reduces to

\begin{eqnarray}\label{gamma1}
\gamma(z) \simeq \frac{\int_0^z \sqrt{\tilde z} \ d{\tilde z}}
{\int_0^z \sqrt{\tilde z} \ d{\tilde z}} = 1.
\end{eqnarray}

In the limit $z \rightarrow \infty$
(very high densities and/or large radii),
from (\ref{hinfty}) and (\ref{finfty}) we get

\begin{eqnarray}\nonumber
\gamma(z) \simeq z^{-\frac{1}{2g}} \cdot 
\frac{\int_0^z \sqrt{\tilde z} \ {\tilde z}^{-1+\frac{1}{2g}} \ d{\tilde z}}
{\int_0^z \sqrt{\tilde z} \ {\tilde z}^{-1} \ d{\tilde z}} =
\frac{\frac{1}{2}}{\frac{1}{2g} + \frac{1}{2}} =
\frac{g}{g+1}, 
\end{eqnarray}
so that asymptotically $\gamma(z) = \frac{1}{2}, \ \frac{2}{3}, \ \frac{5}{6}$
%***
respectively for $g = 1, \ 2, \ 5$ ($\kappa = 1, \ \frac{1}{3}, \ \frac{1}{9}$).
%***
In this limit thus $S(M) \propto M^2$ for $g=1$ and $S(M) \propto M^\beta$
with $1 < \beta < 2$ for $g > 1$. 
Due to (\ref{2m_over_r_far}) or (\ref{mass-temperature})
this means also $S(R) \propto R^2$ for $g=1$ 
and $S(R) \propto R^\beta$ for $g>1$.
Note that this limit can be achieved at any fixed $R$, 
if central energy density $y$
is allowed to increase to sufficiently large values,
and we remain far from the Schwarzschild mass limit.

%***
The graph of $\gamma$ intersects the asymptote
at each extremal point of $\frac{2M}{R}$ of equation (\ref{mass-temperature})
(or of Fig.\ref{mass_vs_T}).
From our discussion about the stability of equilibrium configurations,
this implies that the $z$ region of stability
has both $d\gamma/dz \leq 0$ 
and $\gamma(z)$ above its asymptote.
The intersections with the asymptote 
also mean that for each configuration
corresponding to a local maximum of $\frac{2M}{R}$,
the asymptotic behaviour of $S(M)$ is locally recovered;
for the $g=1$ case this in particular implies that
$S(M) \propto M^2$ for each such finite $z$.
This means that equilibrium configurations exist
(stable, in the case of the first maximum),
with $y$ and $R$ both finite, with exact $S(M) \propto M^2$ 
(and, due to extremality, $S(R) \propto R^2$) local scaling. 
%***

Fig.\ref{s_over_m2} illustrates this.
%%%%%%%%%%%%%%%%%%%%%%%%%%%% figure s_over_m2 %%%%%%%%%%%%%%%%%
\begin{figure}\leavevmode
\begin{center}
\epsfxsize=8cm
\epsfbox{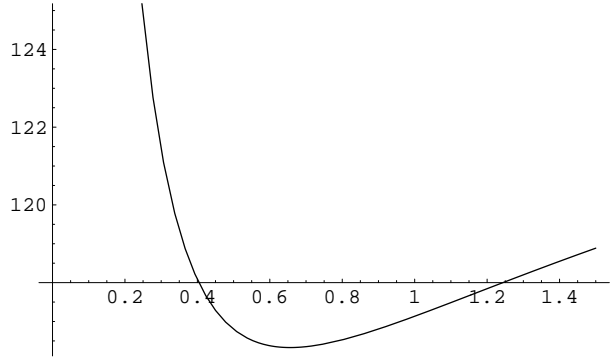}
\caption{$S/M^2$ (arbitrary units) versus $z$ for $\kappa=1$.}
\label{s_over_m2}
\end{center}
\end{figure}
%%%%%%%%%%%%%%%%%%%%%%%%%%%%%%%%%%%%%%%%%%%%%%%%%%%%%%%
We see that for $z=z_{extr}=0.655$ (the first extremal point,
see Fig.\ref{mass_vs_T}), 
$S/M^2$ has a local
extremum, implying $S/M^2 \simeq const$ there.
At the same time this (in addition to (\ref{gamma1}))
verifies the definition of $\gamma$
as given by (\ref{recast}). 

These results confirm and, in some respects, extend what found
for these systems in \cite{banks}.
%***
Within the stability interval in $z$, 
$\gamma$ decreases at increasing $M/R$, as said,
%***
and this is somehow analogous to the result obtained
in \cite{Oppen2}, in the study 
of the relation between $\gamma$ and $M/R$ 
for a self-gravitating system
consisting of a perfect fluid at constant density.
The behaviour of conventional entropy 
$S(R) \propto R^2$ confirms the statement that no horizon
is needed to have an entropy that scales like the area \cite{Oppen2},\cite{Oppen1};
as our systems are far from the Schwarzschild limit (even when $z\rightarrow \infty$)
this statement is confirmed in a strong sense:
to have an area scaling entropy,
not even the approaching to horizon formation 
is required \cite{banks}.

In terms of global temperature $T_0$,
one can show that, 
in the limit $z \rightarrow \infty$,
our systems
have $T_0 \propto 1/R$ \cite{banks},
in addition to the area scaling law for entropy,
so that the analogy with black hole thermodynamics
is quite complete.
This happens also for self-gravitating systems
near to reach their own Schwarzschild radius \cite{Oppen1}.
To see the $T_0$ dependence on $R$,
with reference to equation (\ref{T0}) we write
$T_0 R$ as

\begin{eqnarray}\label{T0R}
T_0 R = b^{-\frac{1}{2g}} y^{\frac{1}{2g} - \frac{1}{2}} 
\frac{z^{1/2}}{[h_1(\sqrt{8\pi z}) f_1(\sqrt{8\pi z})]^{1/2}} \equiv
\eta(y) \xi(z)
\end{eqnarray} 
and
in Fig.\ref{t0_star_r} 
we plot $\xi(z)$ for $g=1,\ 2, \ 5$
(highest: $g=5$, lowest: $g=1$),
together with the constant value
asymptotically approached by $\xi(z)$
in the $g=1$ case revealing the $1/R$
law for $T_0$.
%%%%%%%%%%%%%%%%%%%%%%%%%%%% figure t0_star_r %%%%%%%%%%%%%%%%%
\begin{figure}\leavevmode
\begin{center}
\epsfxsize=8cm
\epsfbox{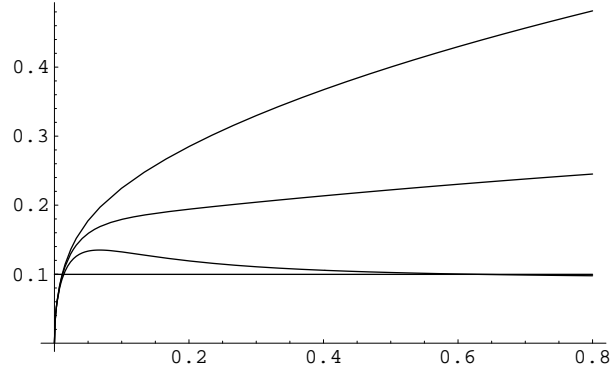}
%***
\caption{$\xi$ (representing $T_0 R$, see text) versus $z$ for 
$\kappa=1, \ \frac{1}{3}, \ \frac{1}{9}$
(highest: $\kappa=\frac{1}{9}$, lowest: $\kappa=1$).}
%***
\label{t0_star_r}
\end{center}
\end{figure}
%%%%%%%%%%%%%%%%%%%%%%%%%%%%%%%%%%%%%%%%%%%%%%%%%%%%%%%
From (\ref{T0R}), together
with equations (\ref{hinfty}) and (\ref{finfty}),
the asymptotic behaviour of $T_0 R$ can analytically be inferred.
We have $T_0 R \propto z^0, \ z^{1/4}, \ z^{2/5}$
(corresponding to $T_0 R \propto R^0, \ R^{1/2}, \ R^{4/5}$)
respectively for $g = 1, \ 2, \ 5$ and
the asymptote in Fig.\ref{t0_star_r} is
at $\frac{1}{2\sqrt{8\pi}}$.
 
For the $g=1$ case note that
every time $z$ is extremal (for $2M/R$),
that happens every time the $\xi(z)$ curve
intersects the asymptote in Fig.\ref{t0_star_r},
the slope is not 0 and thus $T_0 R$ does not scale
as $1/R$ there, not even locally,
notwithstanding the area scaling law for entropy.
This is expected, as
$T_0(z) = \frac{1}{dS_0/dM}$
and
$S_0(M) \propto M^{1/\gamma(z(M))}$, 
so that, 
near each 
%***
extremal $z$, 
%***
the $1/R$ law can set in
only if $\gamma(z) \simeq const$ there,
that is when $z \rightarrow \infty$.
We have thus that only in the limit
$z \rightarrow \infty$ the thermodynamic analogy among
the systems of present study 
and the black holes and the classical systems
of \cite{Oppen1} extends to both $S$ and $T_0$ scaling laws.

Let us consider now the radial profiles of entropy.
We are mainly interested in
their relation with the strength of the gravitational
interaction, as given by the parameter $M/R$, and in the shape they take when
area scaling law for entropy obtains, in particular
in the limiting case $z \rightarrow \infty$ in which
the $1/R$ scaling law for temperature also applies.
From

\begin{eqnarray}\label{define_sigma}
S = 
\int_0^R 4\pi r^2 s(r) \sqrt{h} \ dr =
\int_0^R 4\pi r^2 \sigma(r) \ dr,
\end{eqnarray}
in addition to the entropy local density $s$,
an entropy ``radial'' density $\sigma$ is defined
(in order to compare with \cite{Oppen2}).
As already discussed,
the three components of $s$
as given in (\ref{entropy_local})
have the same radial profiles.
We can then restrict the study
of entropy radial profiles
to the component $s = \rho/T$ alone. 
From (\ref{b}) and (\ref{f1}) 
we have

\begin{eqnarray}\label{sprofile}
s(r) = b^{1/2g} \rho(r)^\frac{2g-1}{2g} =
b^{1/2g} y^\frac{2g-1}{2g} f_1(\sqrt{8\pi y} \ r)^{-g+\frac{1}{2}} =
b^{1/2g} y^\frac{2g-1}{2g} f_1(\sqrt{8\pi z} \ x)^{-g+\frac{1}{2}}
\equiv
\chi(y) \psi_z(x),
\end{eqnarray}

\begin{eqnarray}\label{sigma}
\sigma(r) = b^{1/2g} y^\frac{2g-1}{2g} 
f_1(\sqrt{8\pi z} \ x)^{-g+\frac{1}{2}} h_1(\sqrt{8\pi z} \ x)^\frac{1}{2} \equiv
\chi(y) \omega_z(x),
\end{eqnarray}
where $x \equiv \frac{r}{R}$ covers the interval [0,1]
and $\omega_z(x) = \psi_z(x) h_1(\sqrt{8\pi z} \ x)^\frac{1}{2}$.
Being $f_1$ monotonically increasing, we expect
$\psi_z(x)$ to be monotonically decreasing,
becoming steeper with growing $z$.

For the $g=1$ case, the most interesting one as regards
entropy scaling laws,
Figures \ref{entropy_vs_r} and \ref{entropysigma_vs_r}
report respectively $\psi_z(x)$ and $\omega_z(x)$,
for $z = 0, \ \frac{z_{extr}}{2}, \ z_{extr}=0.655$,
where $z_{extr}$ is the first extremal point,
corresponding to the higher edge 
%***
of the stability interval in $z$. 
%***
The straight line corresponds to $z=0$, while the lowest curve
corresponds to $z = z_{extr}$.
%%%%%%%%%%%%%%%%%%%%%%%%%%%% figure entropy_vs_r %%%%%%%%%%%%%%%%%
\begin{figure}\leavevmode
\begin{center}
\epsfxsize=8cm
\epsfbox{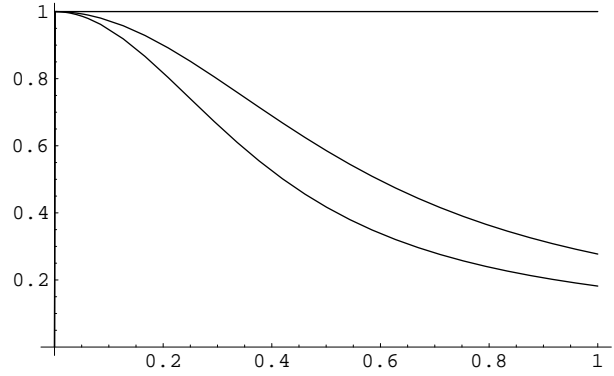}
%***
\caption{Entropy local density $s$, normalized to its value
at the centre $s(0)$, versus radius (normalized to 1) 
for different $z$ values (different $M/R$ values),
$z = 0, \frac{z_{extr}}{2}, z_{extr}$.
The straight line is for $z=0$ ($M/R = 0$);
the steepest curve is for $z=z_{extr}$ ($M/R$ is at its maximum).}
%***
\label{entropy_vs_r}
\end{center}
\end{figure}
%%%%%%%%%%%%%%%%%%%%%%%%%%%%%%%%%%%%%%%%%%%%%%%%%%%%%%%
%%%%%%%%%%%%%%%%%%%%%%%%%%%% figure entropysigma_vs_r %%%%%%%%%%%%%%%%%
\begin{figure}\leavevmode
\begin{center}
\epsfxsize=8cm
\epsfbox{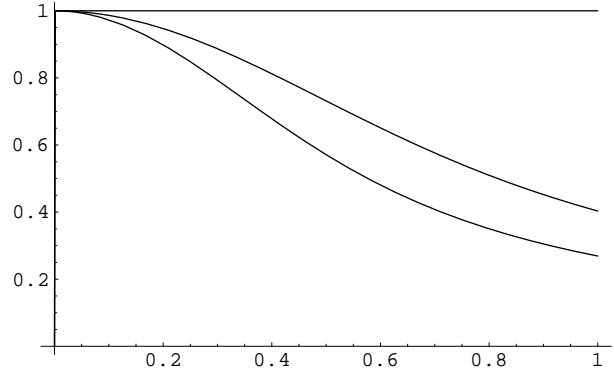}
%***
\caption{Entropy ``radial'' density $\sigma$ (see text), 
normalized to its value at the centre $\sigma(0)$),
versus radius (normalized to 1) 
for the same $z$ values as the previous Figure.
Straight line: $z=0$; lowest curve: $z=z_{extr}$.}
%***
\label{entropysigma_vs_r}
\end{center}
\end{figure}
%%%%%%%%%%%%%%%%%%%%%%%%%%%%%%%%%%%%%%%%%%%%%%%%%%%%%%%
From Fig.\ref{mass_vs_T} (or relation (\ref{mass-temperature}))
we recall that
$M/R$ increases with $z$, from
0 at $z=0$, up to its maximum allowed
value for $z= z_{extr}$.
$s(r)$ or $\sigma(r)$
profiles become thus flat as the strength
of gravitational interaction is negligible,
as expected,
while they become decreasing with $r$
when $M/R \neq 0$, 
being increasingly steeper with $M/R$ growing.
We have thus that the weight of the internal
regions in the determination of total entropy
grows with $M/R$.
This can be seen also in Fig.\ref{entropyslice_vs_r},
where the quantity $x^2 \psi_z(x)$ is reported
($\propto 4\pi r^2 s(r) dr$, the entropy
on the slice at radius $r$).
%%%%%%%%%%%%%%%%%%%%%%%%%%%% figure entropyslice_vs_r %%%%%%%%%%%%%%%%%
\begin{figure}\leavevmode
\begin{center}
\epsfxsize=8cm
\epsfbox{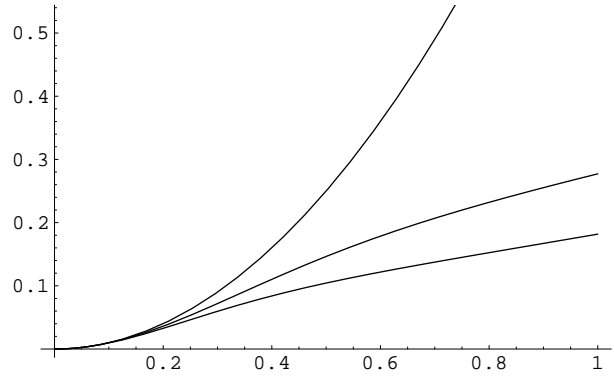}
%***
\caption{$r^2 s$ ($\propto$ entropy of the slice
at radius $r$) versus $r$ with
$s$ normalized to its value at the centre $s(0)$
and $r$ normalized to 1,
for the same $z$ values as previous Figures.
Uppermost curve: $z=0$; lowest curve: $z=z_{extr}$.}
%***
\label{entropyslice_vs_r}
\end{center}
\end{figure}
%%%%%%%%%%%%%%%%%%%%%%%%%%%%%%%%%%%%%%%%%%%%%%%%%%%%%%%
The uppermost curve corresponds to $z=0$, while the lowest
one to $z=z_{extr}$.
The weight of external slices is progressively 
decreasing when $M/R$ grows
from $M/R = 0$ (case with the expected quadratic dependence on $r$
of the entropy of slices of that radius)
to its maximum value.
The behaviour of our system 
is at variance in this respect with
the self-gravitating system consisting
of a perfect fluid with constant density,
studied in \cite{Oppen2}.
In that case the weight of external layers
is shown to increase with $M/R$, as reported by Fig.2
of \cite{Oppen2}, the equivalent of our
Fig.\ref{entropysigma_vs_r}.

%***
For $z>z_{extr}$
the curves for $s(r)$ and $\sigma(r)$
become increasingly steeper and
it is interesting to explore
what is the shape the radial profiles acquire
in the limit $z \rightarrow \infty$;
as we know in fact in this limit
for the case $g=1$ we have both the area scaling
law for entropy and the $1/R$ law for temperature.
For $z\rightarrow\infty$,
from (\ref{sprofile}) and (\ref{finfty})
we find
%***

\begin{eqnarray}\label{slimit}
\psi_z(x) \simeq \left[\left(\frac{2m}{r}\right)_\infty^{-\frac{1}{g}} (8\pi z)^\frac{1}{g}
x^\frac{2}{g} \right]^{-g+\frac{1}{2}} \propto
x^{-2+\frac{1}{g}}
\end{eqnarray}
and from (\ref{hinfty})
a same behaviour for $\omega_z(x)$.
This means that in this limit
$s(r)$ and $\sigma(r)$ 
%***
behave
%***
like $1/r$ if $g=1$ and like $1/r^\delta$
with $1 < \delta < 2$ for any other allowed $g$;
in the limit $g\rightarrow\infty$ (dust),
$\delta=2$.
We see thus that for our systems,
when we are at conditions in which
both the area scaling for entropy and
the $1/R$ law for temperature set in,
the entropy radial profile goes as $1/r$,
with a larger weight from internal layers
in the determination of total entropy,
as compared to the no-gravity case.
This is at variance with what happens
for self-gravitating
classical systems that reach the above scaling laws
for entropy and temperature while approaching
their Schwarzschild radius, 
for which
it is demonstrated that
in this limit all entropy lies on the surface \cite{Oppen1}.

\section{Conclusions}

In conclusion we have seen that the area scaling law for entropy
as well the $1/R$ law for temperature (the celebrated properties
of black hole thermodynamics) can set in also for classical finite systems
even at circumstances very far from the formation of an horizon.
We have shown moreover that when this happens, entropy radial profile
may be completely different from what found for the systems
that gain the above scaling laws while approaching horizon formation,
in which case entropy lies on their surface \cite{Oppen1}.

We have studied the onset of these properties for systems consisting
of perfect fluids with linear equation of state $p = \kappa \rho$,
constrained in a finite volume by an external pressure.
The above properties set in at limiting conditions of causality \cite{banks} ($\kappa =1$) 
and size/energy density ($R$ and/or $\rho\rightarrow\infty$), 
being however safely far from horizon formation limit. 
These results have been obtained generalizing the methods introduced
in \cite{Oppen2} 
for the study of non-extensivity of thermodynamical parameters
in gravitating systems,
to the case in which an external pressure is applied.
The entropy radial profile we find in the limiting circumstances
in which the mentioned scaling laws for entropy and temperature set in
is $s(r) \propto 1/r$, so that a heavy contribution is
given by the internal layers, contrary to what happens for the systems \cite{Oppen1}. 
This is thus a case in which 
the whole volume contributes to
the total number of degrees of freedom of the system
(with a heavier specific contribution from internal regions), 
being however this number scaling like the area.

\end{document}